\documentclass[prl,twocolumn,showpacs,superscriptaddress,groupedaddress]{revtex4}
\usepackage{amssymb}
\usepackage{color,graphicx}
\usepackage{amsmath}
\usepackage{amsbsy}
\usepackage{amsthm}
\usepackage{bbm}
\usepackage{bm}
\usepackage{epsfig}
\usepackage{float}
\usepackage{xcolor}
\usepackage{subfigure}
\usepackage{comment}

\begin{document}
\author{Marko Toro\v{s}}
\email{marko.toros@ts.infn.it}

\author{Giulio Gasbarri}
\email{giulio.gasbarri@ts.infn.it}

\author{Angelo Bassi}
\email{bassi@ts.infn.it}

\affiliation{Department of Physics, University of Trieste, 34151 Miramare-Trieste, Italy \\
Istituto Nazionale di Fisica Nucleare, Sezione di Trieste, Via Valerio 2, 34127 Trieste, Italy}
\title{Colored and Dissipative Continuous Spontaneous Localization model
and Bounds from Matter-Wave Interferometry}
\date{\today}
\begin{abstract}
Matter-wave interferometry is a direct test of the quantum superposition principle for massive systems, and of collapse models. 
Here we show that the bounds placed by matter-wave interferometry depend weakly on the details of the collapse mechanism. Specifically, we compute the bounds on the CSL model and its variants, provided by the the KDTL interferometry experiment of Arndt's group [Phys. Chem. Chem. Phys., 2013, 15, 14696-14700], which currently holds the record of largest mass in interferometry. 

 We also show that the CSL family of models emerges naturally by considering a minimal set of assumptions. In particular, we construct the dynamical map for the colored and dissipative Continuous Spontaneous Localization (cdCSL) model, which reduces to the CSL model and variants in the appropriate limits. In addition, we discuss the measure of macroscopicity based on the cdCSL model.
\end{abstract}
\maketitle

Since the discussion of Schr\"odinger on the consequences of  the quantum superposition principle when applied to macroscopic objects \cite{Schrodinger},  the debate about the emergence of the classical world from quantum physics has not been resolved. Why do we not see macroscopic superpositions?

This question was confined to a speculative debate for a very long time. Nowadays the impressive technological progress has brought it into  the realm of experimental physics. Matter-wave interferometry started with single particles \cite{Jonsson} and now involves large molecules with up to $10^4$ a.m.u. \cite{C3CP51500A}. Optomechanics promises to superimpose much larger masses \cite{PhysRevLett.107.020405}. How big has the system to be, to represent a significant test of the quantum superposition principle?

Collapse models \cite{collapse_review1,collapse_review2} offer a quantitative answer. They have been proposed to explain the quantum-to-classical transition through nonlinear and stochastic modifications of the Schr\"odinger equation. These modifications have a negligible effect on the dynamics of microscopic systems, like atoms and small molecules. At the same time, when atoms and molecules glue together to form more complex systems, the collapse process is amplified, to the point that macroscopic objects are always well localized in space.  

Collapse models are phenomenological models, so without much surprise several models have been proposed over the years. They all have the same structure, and  general arguments show that this has to be the case \cite{trace}, to avoid a conflict with relativity. But they differ, sometimes significantly, in the details. The Ghirardi-Rimini-Weber (GRW) model \cite{GRW} was the first proposed in the literature, and was soon after generalized in the Continuous Spontaneous Localization (CSL) model \cite{CSL1,CSL2} to include identical particles into the description. This has now become the reference model. 

The CSL model contains two phenomenological parameters: the correlation length $r_C$ of the noise, which defines the spatial resolution of the collapse, and a  rate $\lambda$, which sets the strength of the collapse process. Originally, the following values were suggested:  $\lambda \simeq 10^{-16}$ s$^{-1}$ for $r_C \simeq 100\text{nm}$ \cite{GRW}. (According to \cite{CSL1,CSL2}  one has $\lambda \simeq 10^{-17}\text{s}^{-1}$ for  $r_C \simeq 100\text{nm}$ .) More recently, Adler suggested a much stronger value for $\lambda$ ($\simeq 10^{-8\pm2}$ s$^{-1}$ for $r_C \simeq 100\text{nm}$  and $\simeq 10^{-6\pm2}$ s$^{-1}$ for $r_C \simeq 1$ $\mu\text{m}$) \cite{1751-8121-40-12-S03}.

The CSL model, like the GRW model, violates the energy conservation principle, as the noise driving the collapse induces a Brownian motion, increasing the kinetic energy. This feature has been exploited to devise non-interferometric tests of collapse models \cite{PhysRevLett.112.210404,PhysRevA.89.032127,PhysRevLett.113.020405,PhysRevLett.114.050403}, which so far  place the strongest bounds on the collapse parameters \cite{2015arXiv151005791V,Curceanu}, ruling out Adler's values by some two orders of magnitude.

The violation of  energy conservation  can be tolerated in a phenomenological model, but eventually has to be removed. This has been partially achieved by introducing the dissipative CSL (dCSL) model \cite{Smirne:2014paa}. It behaves like the CSL model as far as the collapse process is concerned. At the same time, the energy does not steadily increase, but reaches an asymptotic finite value, controlled by a new parameter $T$, which plays the role of the temperature of the noise. If the collapse of the wave function is a universal feature, then the noise is spread over the universe, and much likely has a cosmological origin. Therefore, a reasonable value for its temperature is $T \sim 0.1-10 $K.

In all these models the noise is assumed to be white. This is very convenient from the mathematical point of view, but is not physical. Real noises always have a non-flat spectrum. A CSL model with a colored noise (cCSL) has been introduced \cite{PhysRevA.48.913,1751-8121-40-50-012,1751-8121-41-39-395308}. A new  parameter appears, the cut off frequency $\Omega$.  Also in this case, the collapse properties are preserved, but the Brownian motion induced on quantum systems changes significantly at high frequencies. If the noise has a cosmological origin, a reasonable value for the cut off is $\Omega \sim 10^{10}-10^{11}$Hz, as that of some of the most common cosmological backgrounds \cite{0295-5075-92-5-50006}.

The existence of all these models poses a problem: how can they all be tested? Non-interferometric tests, such as those proposed in \cite{PhysRevLett.112.210404,PhysRevA.89.032127,PhysRevLett.113.020405}, might soon rule out the CSL model. This will be a significant result. But it is not clear whether they will rule out also the dCSL and/or cCSL models.

Here we show that matter-wave interferometry, being a direct test of the quantum superposition principle, is quite insensitive to variantions of the CLS model. At the same time, we explore the region of parameter space excluded by existing matter-wave experiments, which was only partially analyzed in the past \cite{1751-8121-45-6-065304, PhysRevA.83.043621}. In the end, we will present a comprehensive picture of how such experiments constrain the CSL model and its variations so far. These bounds are weaker than those placed by non-interferometric tests, but robust. Detailed calculations for each considered collapse model are presented in \cite{matterwave}. 

As part of the analysis, we show that the family of CSL models here discussed naturally emerges by imposing Galilei space-time symmetries on the dynamics, driven by Gaussian operators. Specifically, we construct the colored and dissipative Continuous spontaneous localization (cdCSL) model, that reduces to the cCSL and dCSL models in the appropriate limits. We conclude by briefly discussing a macroscopicity of matter-wave interferometry experiments.

\begin{figure}[!t]
\begin{center}
\includegraphics[width=1.0\columnwidth,trim={50 140 80 110},clip]{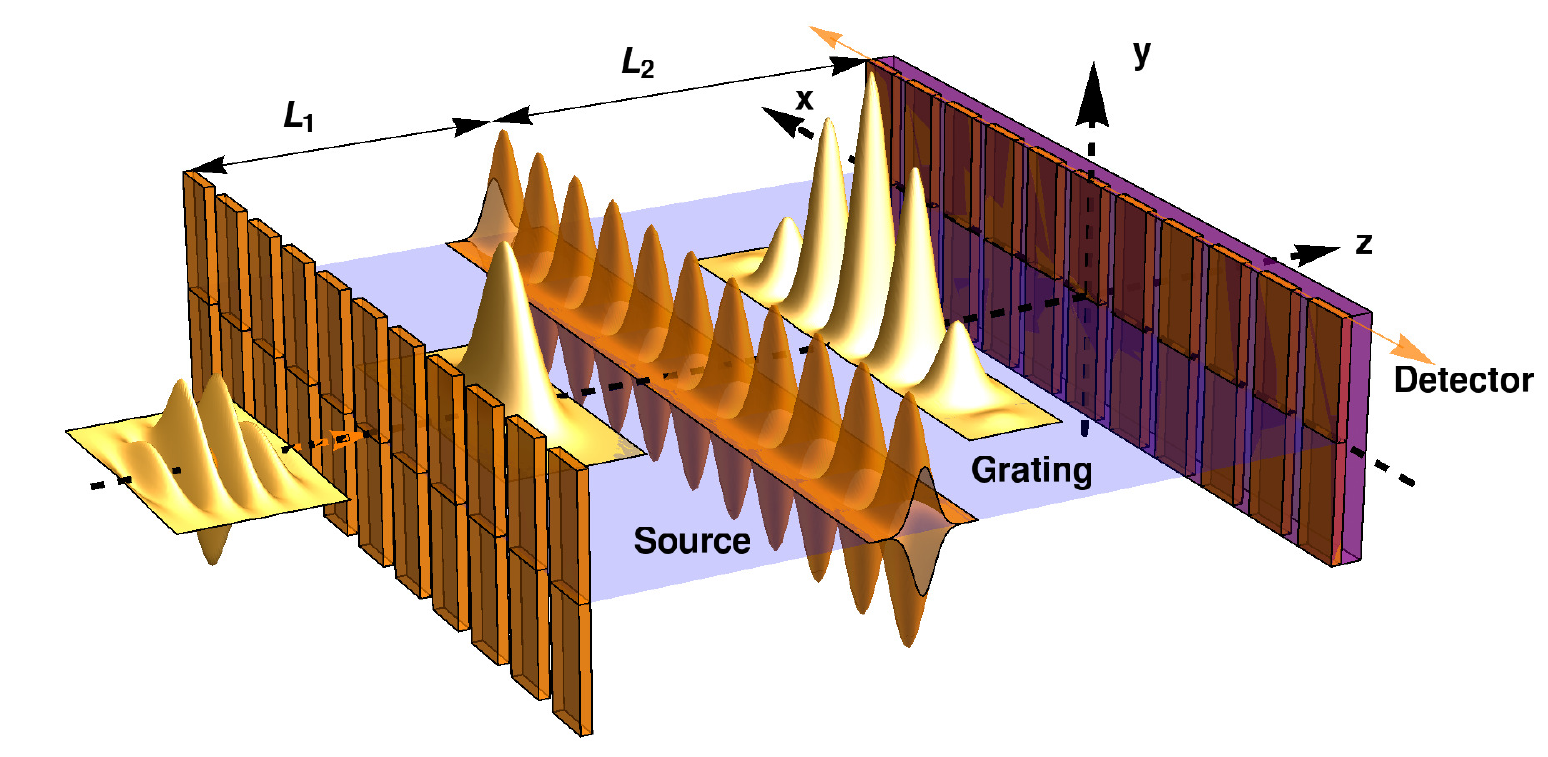}
\caption{A molecular beam from an incoherent source propagates along the $z$ axis. Each molecule, individually, propagates to an optical grating produced by a standing light wave, where its wave function is diffracted and subsequently recorded by a detector. The molecules, individually recorded, gradually form an interference pattern. The distance (flight time) from the source to the grating is $L_1$ ($t_1$) and the distance (flight time) from the grating to detector is $L_2$ ($t_2$). In the KDTL experimental setup, there are two additional mechanical gratings blocking part of the molecules: the mechanical grating located immediately after the source is held fixed, and prepares the beam for diffraction. The mechanical grating immediately before the detector  moves along the $x$ axis. The detector records molecules that arrive at all points along the $x$ axis in a certain amount of time (for a given displacement of the third grating from its original position).}
\label{Fig:1}
\end{center}
\end{figure}

\noindent{\it Theoretical analysis -- }
We consider the Kapitza-Dirac-Talbot-Lau (KDTL) interferometer schematically depicted in Fig.~\ref{Fig:1}, which holds the world record for the largest mass employed ($10^4$ a.m.u.).  
The dynamics for the density matrix describing the motion of the center of mass of a rigid body along the $x$ direction, while propagating towards the grating along the $z$ direction, has a similar structure for all collapse models. Its solution in the paraxial approximation~\cite{hornberger2009theory}, using the characteristic function approach~\cite{savage1985damping,smirne2010quantum}, can be expressed as follows~\cite{GRW,smirne2014dissipative,matterwave}:
\begin{equation} \label{eq:1}
\begin{split}
\rho(x,x',t)  = & \frac{1}{2\pi\hbar} \int_{-\infty}^{+ \infty}  dk \int_{-\infty}^{+ \infty} dw\,  e^{-ikw/\hbar} F(k, x-x',t)  \\
 & \times \rho^{\text{\tiny QM}}(x+w, x'+w, t),
\end{split} 
\end{equation}
where $\rho^{\text{\tiny QM}}$ encodes the standard free quantum evolution, and $F$, which depends on $\lambda$ and $r_C$, the effect of the collapse. When $F = 1$ we have the standard quantum behavior. Different functions $F$ are associated to different collapse models  (see Appendix A1 for details). Eq.~\eqref{eq:1} predicts the following  pattern (position distribution of the molecules) at the detector, corresponding to what is actually measured~\cite{PhysRevA.70.053608, matterwave}: 
\begin{equation} \label{eq:2}
S(x) = \sum_{n = -\infty}^{\infty} A^*_n C^*_n B_n D\left( \frac{2 \pi n}{d} \frac{L}{k} \right) e^{2 \pi i n x/ d},
\end{equation}
where $d$ is periodicity of the optical elements, $L=L_1=L_2$ (see Fig.~\ref{Fig:1}) and $k$ is wave number of the matter wave. The coefficients $A_n$, $B_n$ and $C_n$ are geometric factors related to the Fourier transform of the transmission functions associated to the three optical elements (Source, Grating and Detector respectively) and encode their effect on the beam~\cite{1464-4266-5-2-362}. The function $D(x)$, which contains the dynamical information about the collapse effect during the propagation of the beam, is related to the $F$ function~\cite{PhysRevA.83.043621,matterwave}: 
\begin{equation}\label{eq:3}
D(x) = F(-\hbar kx/ L_2,0,t_2) F(\hbar kx /L_1,x,t_1), 
\end{equation}
where the first and second factor on the right-hand side describe the amount of deviation from quantum mechanics that accumulates during the flight from source to grating and from grating to detector, respectively. Specifically, the function $D$ inherits the dependence on $\lambda$ and $r_C$ from  $F$. When $D = 1$ we have again the standard quantum behaviour.  Fig.~\ref{Fig:2} shows the values of $D$ for all collapse models considered here.

\noindent{\it The amplification mechanism -- }
Matter-wave interferometry creates the superposition of different center-of-mass spatial states of a macro-molecule, which eventually interfere with each other. From the theoretical point of view, under the {\it rigid-body} approximation the molecule can be treated as a single particle satisfying the collapse dynamics as given by Eq.~\eqref{eq:1}. In this case, the collapse rate $\lambda$ for a single nucleon has to be replaced by  a rate $\Lambda$ associated to the center of mass,  which is a function of $\lambda$, enhanced by a geometric factor depending on the geometry and number of nucleons in the molecule. This is the mathematical description of the amplification mechanism~\cite{GRW, collapse_review1,collapse_review2}.  

For a rigid body, when the wave function of the molecule is delocalized {\it more} than its size, as it is the case for the experiment under consideration, a reasonable expression for $\Lambda$ is \cite{1751-8121-40-12-S03,matterwave}:
\begin{equation}\label{eq:4}
\Lambda =  \frac{n_A}{n(r_C)} \left( \frac{m_A n(r_C)}{m_0}  \right)^2 \lambda,
\end{equation} 
where $n(r_C)$ is the number of atoms (nuclei) contained in a volume of linear size $r_C$, while $m_A$ is the atomic mass, $n_A$ is the number of atoms and $m_0$ is the proton reference mass. 

\begin{figure}[!t]
\begin{center}
\includegraphics[width=1.0\columnwidth]{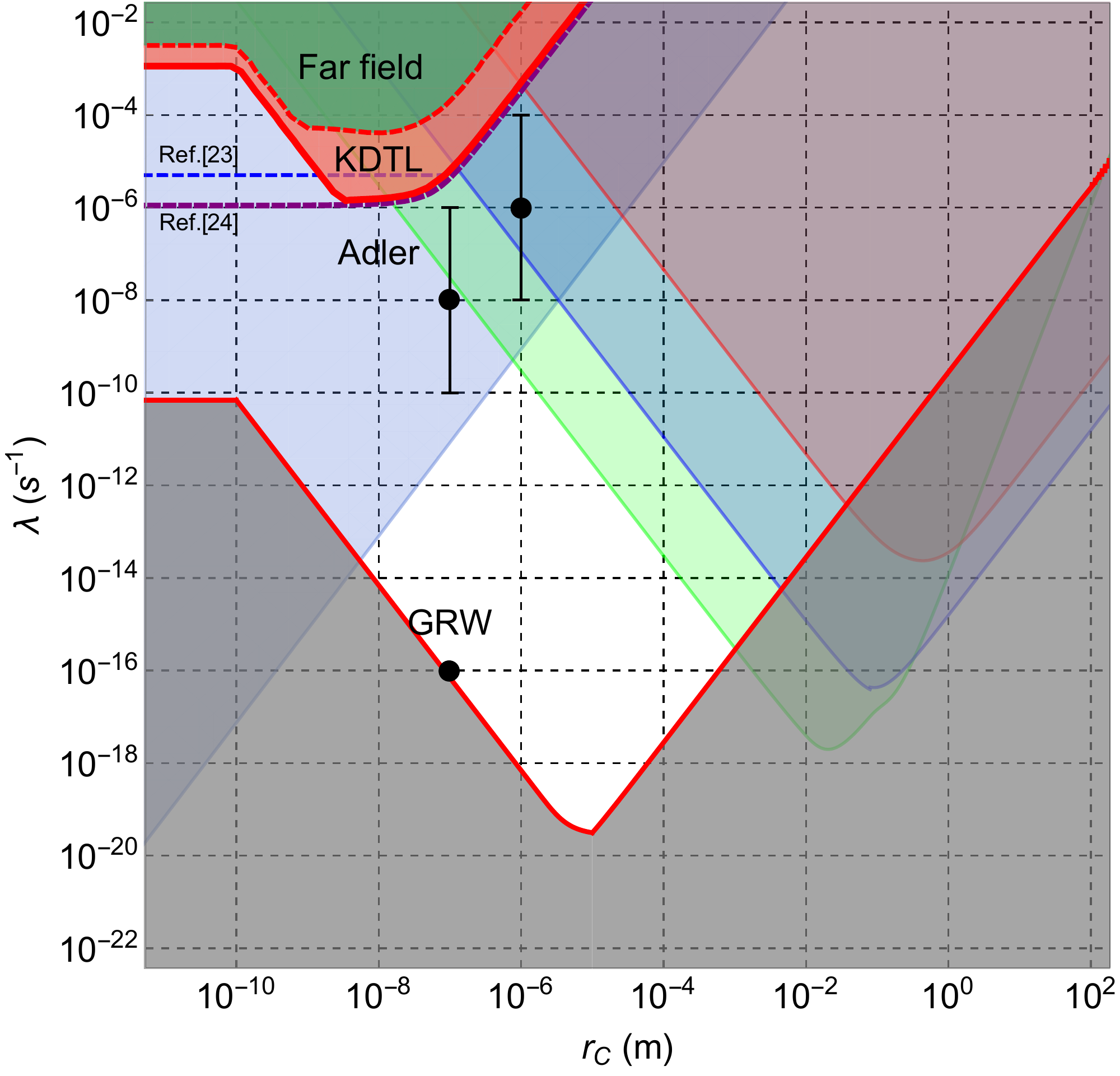}
\caption{
Parameter diagram for the CSL, dCSL and cCSL models. The exclusion zone, given by the gray shaded zone at the bottom (bordered by the red solid line), arises from the requirement that collapse models become effective for macroscopic system. The red shaded zone at the top corresponds to the upper bounds set by the KDTL~\cite{C3CP51500A} experiment discussed in the text. We have also reported the bounds from the far field experiment~\cite{Real, MSCH}, given by the the dark green exclusion zone, which are roughly 2 orders of magnitude weaker. For comparison we have included the bounds from X-ray experiments \cite{Curceanu}, valid for the CSL model and the cCSL model with frequency cutoff $\Omega\gg10^{18}Hz$, given by the light blue exclusion zone on the left, and the bounds from  LIGO, LISA Pathfinder and AURIGA \cite{carlesso2016experimental}, analyzed so far for the CSL model only, given by the exclusion zones on the right, shaded in light blue, light green and light red, respectively.
We have also included for reference, the GRW~\cite{GRW} values $(\lambda=10^{-16}\text{s}^{-1},r_{C}=10^{-7}\text{m})$ and the values proposed by Adler \cite{1751-8121-40-12-S03}:  $(\lambda=10^{-8\pm2} \text{s}^{-1},r_{C}=10^{-7}\text{m})$ and $(\lambda=10^{-6\pm2}\text{s}^{-1},r_{C}=10^{-6}{m})$. The dashed blue and purple lines denote the KDTL bounds estimated using the analysis from~\cite{1751-8121-45-6-065304} and \cite{PhysRevA.83.043621}, respectively. We note that for values of $r_C$ smaller than the size of the macro-molecule ($\simeq 10^{-8}m$), the bounds on $\lambda$ become less stringent.}
\label{Fig:4}
\end{center}
\end{figure}

\noindent{\it The interference pattern -- }
Collapse models predict a loss of visibility, with respect to standard quantum mechanics. This effect can be used to set an upper bound on the collapse parameters, and exclude a region of parameter space, where the parameters take too strong values. Since we are interested in the order of magnitude, a $\chi^2$ minimization procedure to compare the theoretical predictions, computed using Eqs.~\eqref{eq:2},~\eqref{eq:3} and~\eqref{eq:4}, to the experimental data will suffice. The outcome is reported in Fig.~\ref{Fig:4}. 

The plot depicts two exclusion zones. The one at the bottom comes from the requirement that the model localizes macroscopic objects fast enough (shaded gray zone). Specifically, using Eqs.~\eqref{eq:1}, one imposes that the off-diagonal elements of the density matrix $\rho(x,x',t)$ are supressed fast enough. If this does not happen, then the model fails to satisfy the fundamental requirement for which it was first formulated. To be quantitative, we required that a single-layered graphene disk of radius $\simeq 0.01$~mm (minimum resolution of the human eye) is localized within $\simeq 10$~ms (perception time of the human eye). The plot shows that according to our classicality criterion the original GRW value for $\lambda$ is  the lowest possible value (for $r_C \simeq 100$~nm) for collapse models to explain classicality. Clearly, this lower bound can be shifted also by several orders of magnitude, depending on the chosen criterion for classicality~\cite{1751-8121-45-6-065304}.

The exclusion zone at the top comes from comparison with the KDTL experiment in \cite{C3CP51500A} (shaded red  zone). First we have considered the standard CSL model, which depends only on $\lambda$ and $r_C$. The exclusion zone is identified by the red line in Fig.~\ref{Fig:4}. The border of the exclusion zone highly depends on the shape and size of the molecule through the amplification mechanics given in Eq.~\eqref{eq:4}. In particular, the slope changes significantly from $r_C=10^{-10}\text{m}$ (comparable to the atomic radius) to $r_C=10^{-8}\text{m}$ (comparable to the molecular radius). The slope of the lower bound instead changes at $r_C=10^{-5}m$ (the radius of the disk).

Next, we considered the dCSL model. Besides $\lambda$ and $r_C$, it depends also on the temperature $T$ of the collapse field and on the average noise field velocity parameter $\bm{u}=(u_x,u_y,u_z)$. These new parameters can be understood by looking at the quantum linear Boltzmann equation~\cite{Vacchini200971}, which has the same mathematical form as the dCSL master equation, and describes the motion of a particle (system) immersed in a bath of particles (noise field) of temperature $T$ moving with average velocity $\bm{u}$. The exclusion zone coincides with the CSL exclusion zone for a large set of temperatures and velocities. Only when we consider very strong dissipation (e.g. $T = 10^{-12}K$) or relativistic velocities (e.g. $u_x = 10^8\text{ms}^{-1}$), the dCSL exclusion zone becomes noticeably different from the CSL exclusion zone (see Appendix A2 for details). 

Finally, we considered the cCSL model, which depends on the cutoff frequency $\Omega$, in addition to $\lambda$ and $r_C$. Our analysis applies to cCSL with $\Omega \gg 10^{13}\text{Hz}$, for which the exclusion zone coincides with the white noise CSL exclusion zone denoted by the red lines (see Appendix A2 for details). 

These bounds have to be compared with the non-interferometric bounds in the literature, which currently give stronger bounds on the CSL model. While interferometric experiment test the main feature of all collapse models, i.e. the suppression of macroscopic superpositions, non-interferometric experiments test secondary features of the dynamics given in Eq.~\eqref{eq:1} for a specific collapse model. The experimental absence of these secondary effects can then be used to set bounds on the parameters of the tested dynamics. We now discuss the most relevant non-interferometric bounds.

The strongest one comes from experiments on spontaneous X-ray emission \cite{Curceanu} (light blue shaded zone on the left). Loosely speaking, the coupling with the noise field induces a random accelerated motion of a charged particle, which is then expected to radiate. The absence of spontanteously emitted X-rays sets a bound that is several orders of magnitude stronger than that coming from matter-wave interferometry. However, as argued in \cite{Donadi201470} it can be evaded by considering a cCSL model with a frequency cut off as high as $10^{18}$ Hz, which is much higher than what reasonably expected. It is not clear yet what happens in the case of the dCSL model. 

Another strong bound comes from the collapse induced brownian motion (light green shaded zone on the right), which has been searched for, by analyzing the noise spectrum of LISA Pathfinder~\cite{carlesso2016experimental}. For completeness, we have included also the bounds obtain from LIGO (light blue shaded zone on the right) and AURIGA (light red shaded zone on the right). In a nutshell, the coupling with the noise field is expected to induce a small random motion even for a macroscopic object. By precisely monitoring the position of the object one can set strong bound on the collapse parameters. At present, it is not clear how these bounds are affected by the inclusion of dissipation (dCSL model) and color (cCSL model). 

Recently, the cCSL and dCSL bounds from cold atom experiments~\cite{bilardello2016bounds} have been fully analyzed. The bounds are obtained by looking for a possible anomalous heating (or cooling) of a gas of cold atoms, which is due the collapse dynamics. Specifically, the cCSL bounds from cold atom experiments are comparable with the bounds from spontaneous X-ray emission experiments \cite{Curceanu} (light blue shaded zone on the left) and do not change for frequency cutoffs higher than $10^{6}$ Hz. On the other hand, the dCSL bounds from cold atom experiments change significanly when dissipation is included in the analysis, as shown in Fig.~(8) of~\cite{bilardello2016bounds}.

\noindent{\it Colored and dissipative CSL-- } Thus far we have discussed the bounds on the parameters of the CSL model and its variants which have been proposed in the literature. However, at this level it is not clear to which extent such variants are arbitrary. Here we show that this arbitrariness is limited.
Limiting the discussion to the framework of Gaussian maps~\cite{diosi2014general}, which describes a very broad class of dynamics, we show that the "CSL family" of models (including CSL and the variants considered here above) emerges naturally by considering a set of minimal assumptions: (i) translational covariance, (ii) stationary Gibbs state and (iii) Gaussian operators. This result gives to the bounds from matter-wave interferometry in a different perspective: they describe the experimental bounds on the non-unitary modifications of the quantum dynamics satisfying assumptions (i)-(iii).

Assumptions (i) and (ii) are both related to Galilei space-time symmetries. Specifically, assumption (i) implies the following dynamical map (in the interaction picture)~\cite{gasbarri2017general}:
\begin{align}
\mathcal{M}_{t}=&\mathcal{T}\exp\Big\{\int_{0}^{t}d\tau\int_{0}^{t}ds\int_{\mathbb{R}^3} d \bm{Q} 
\, G(\tau-s)\nonumber\\
&\big(  [e^{\frac{i}{\hbar}\bm{Q}\cdot\hat{\bm{x}}_{L}(s)}J_{L}(\hat{\bm{p}},\bm{Q})]\,[e^{-\frac{i}{\hbar}\bm{Q}\cdot\hat{\bm{x}}_{R}(\tau)}J_{R}^{\dagger}(\hat{\bm{p}},\bm{Q})]\nonumber \\
 & -\theta_{\tau,s}[J_{L}^{\dagger}(\hat{\bm{p}},\bm{Q})e^{-\frac{i}{\hbar}\bm{Q}\cdot\hat{\bm{x}}_{L}(\tau)}]\,[e^{\frac{i}{\hbar}\bm{Q}\hat{\bm{x}}_{L}(s)}J_{L}(\hat{\bm{p}},\bm{Q})]\nonumber \\
 & -\theta_{s,\tau}[J_{R}(\hat{\bm{p}},\bm{Q})e^{\frac{i}{\hbar}\bm{Q}\cdot\hat{\bm{x}}_{R}(s)}]\,[e^{-\frac{i}{\hbar}\bm{Q}\cdot\hat{\bm{x}}_{R}(\tau)}J_{R}^{\dagger}(\hat{\bm{p}},\bm{Q})]\Big\}
,\label{eq:translationC1}
\end{align}
where $G$ is a temporal correlation function with correlation time $\tau_{C}$ and $L$ ($R$) denotes operators acting on the statistical operator from the left (right). For simplicity one can choose an exponential correlation function $G(\tau-s)=\frac{1}{2\tau_C}e^{-\frac{|\tau-s|}{\tau_C}}$. Setting $G(\tau-s)=\delta(\tau-s)$, i.e. $\tau_C \rightarrow 0$, one recovers the translationally covariant (Markovian) Lindblad master equation in Appendix~A3.

The only remaining freedom in Eq.~\eqref{eq:translationC1} is in the choice of the operators $J$. To remove it, one is tempted to impose the full Galilei symmetry group, specifically covariance under boosts, but this leads to an infinite temperature increase for an isolated system~\cite{gasbarri2017general}. To avoid this unphysical feature, we relax this assumption, and require instead that an isolated system has the (ii) stationary Gibbs state:
\begin{equation}
\hat{\rho}_{\text{asm}}=\left(\frac{\beta}{2m\pi}\right)^{3/2}\exp(-\beta\hat{H}),\label{eq:rho_asm}
\end{equation}
where $\hat{H}=\frac{\hat{\bm{p}}^{2}}{2m}$, $\hat{\bm{p}}$ is the center of mass momentum operator, $m$ is the mass of the system, $\beta=1/(k_{B}T)$, $k_{B}$ is the Boltzmann constant and $T$ is the thermalization temperature. Assumption (ii), together with the assumption that $J$ is (iii) Gaussian in $\hat{\bm{p}}$ and $\bm{Q}$, leads us to consider:
\begin{equation}
J(\hat{\bm{p}},\bm{Q})=\sqrt{\lambda\frac{m^{2}}{m_{0}^{2}}\left(\frac{r_{C}}{\sqrt{\pi}\hbar}\right)^{3}}e^{-\frac{r_{C}^{2}}{2\hbar^{2}}((1+k_{T})\bm{Q}+2k_{T}\hat{\bm{p}})^{2}},
\label{eq:JoperatorRe}
\end{equation}
where  $k_T=\frac{\hbar^2}{8 m r_C^2 k_B T}$ and the overall normalization has been chosen to match standard CSL notation (see Appendix~A3 for more details).  

Equation~\eqref{eq:translationC1} and~\eqref{eq:JoperatorRe} define the dynamical map of the colored and dissipative Continuous Spontaneous Localization model (cdCSL) which generalizes and embodies all previous models. On the one hand, Eq.~\eqref{eq:translationC1} reduces to the dCSL dynamics in the Markovian limit ($\tau_{C}\rightarrow0$) while, on the other hand, it reduces to the cCSL map in the non-dissipative limit ($T\rightarrow\infty$). The corresponding cdCSL dynamics for state vectors (given by a non-Markovian stochastic differential equation) can be written with the formalism developed in~\cite{1751-8121-40-50-012,1751-8121-41-39-395308}.

\noindent{\it Macroscopicity measure -- }
It is interesting to ask how much a given experiment explores the boundary between the quantum and classical regime. To give a quantitative answer, one can consider a classicalization map, i.e. a non unitary dynamics that replaces the quantum mechanical evolution, which can be used to define an index of classicality, i.e. a macroscopicity measure~\cite{nimmrichter2013macroscopicity,nimmrichter2014macroscopic}. It becomes rather natural to consider the cdCSL map as classicalization map and define the macroscopicity measure as~\cite{gasbarri2017general}:
\begin{equation}
\mu=\log (\tau/1\,\text{s})+\log ((m_0/m_e)^2),
\label{macroscopicity}
\end{equation}
where $\tau=\lambda_{\text{min}}^{-1}$, and $\lambda_{\text{min}}$ is the minimum excluded value of $\lambda$ for a given experiment; the constant offset $\log ((m_0/m_e)^2)$, where $m_e$, $m_0$ are the electron and proton mass, respectively, is introduced for consistency with the measure introduced in~\cite{nimmrichter2013macroscopicity}.  

The measure $\mu$ can be interpreted as follows: the stronger the bounds on $\lambda$ the more the experiment pushes towards the quantum-to-classical boundary. For the KDTL matter-wave interferometry one can read $\lambda_{\text{min}}\approx 10^{-6}\text{s}^{-1}$ from Fig.~\ref{Fig:4} and, using Eq.~\eqref{macroscopicity}, one obtains the value $\mu \approx 12.5$. For comparison, an experiment probing the GRW value $\lambda=10^{-16} \text{s}^{-1}$ would achieve a  macroscopicity $\mu \approx 22.5$. 
 
In general however, the value of $\lambda_{\text{min}}$ and thus of $\mu$ will depend on the values of the cdCSL parameters $\tau_C,T,u_x$, which quantify the degree of non-Markovianity and dissipation, i.e. $\mu=\mu(\tau_C,T,u_x)$.  We have shown that the bounds from matter-wave interferometry depend weakly on $\tau_C,T,u_x$ for a large range of values (see Appendix A2 for more details). For this range of values, the measures introduced in~\cite{gasbarri2017general} and~\cite{nimmrichter2013macroscopicity}  then coincide. In other words, our analysis  sets the limits of validity of the macroscopicty measure introduced in~\cite{nimmrichter2013macroscopicity}. For other experiments, where dissipative and non-Markovian effects might be important, the range of $\tau_C,T,u_x$ values, where the two measures agree, might be significantly narrower~\cite{gasbarri2017general}. In this sense, matter-wave interferometry, unlike other indirect tests of the superposition principle, provides a  better, more robust probe for the quantum-to-classical transition. 

To discuss the macroscopicity of future matter-wave interferometry experiments it is useful to obtain an approximate formula from Eq.~\eqref{macroscopicity}. This can be done in the following way. The collapse mechanism becomes relevant, when time of flight $t$ and the effective collapse rate $\Lambda$ satisfy the condition $\Lambda t\simeq 1$~\cite{1751-8121-45-6-065304}. In particular, the minimal value $\lambda_{\text{min}}=1/(t (m/m_0)^2)$ that satisfies this condition is found by setting $n(r_C)=n_A$ in Eq.~\eqref{eq:4}. Thus inserting this expression of $\lambda_{\text{min}}$ into Eq.~\eqref{macroscopicity} we obtain an approximate formula~\cite{nimmrichter2013macroscopicity}:
\begin{equation}
\mu=\log \left(\frac{t}{1\,\text{s}}\right)
+2 \log \left(\frac{m}{m_0}\right)
-\log \left(|\text{ln} (f)|\right),
\label{mu}
\end{equation}
where the last term has been added to account for the measurement apparatus and $f$ denotes the fraction of the expected visibility we can measure with confidence. 

From Eq.~\eqref{mu} we see that the  macroscopicity of matter-wave interferometry experiments scales logarithmically with both the time of flight $t$ and the macromolecular mass $m$. This makes the task of achieving significantly higher macroscopicties a non-trivial task. Specifically, for Earth-bound experiments, the free fall of macromolecules poses severe limitations for future experiments~\cite{nimmrichter2011testing}. Thus, to achieve a significantly higher macroscopicity in matter-wave inteferometry one might need to consider experiments in space~\cite{kaltenbaek2016macroscopic}.

\noindent{\it Conclusion -- }
In summary, matter-wave interferometry is difficult to implement, as it is difficult to create macroscopic superpositions of massive objects. But it represents a direct test of the quantum linearity. Because of this, as proven here, it allows to test the CSL model and all its variations, and place  bounds, which depend weakly on the color and dissipation of the model, and therefore give a strong indication of which scales (size and mass of the system) the quantum superposition principle is valid.  

\noindent{\it Acknowledgements -- }
The authors acknowledge financial support from the EU project NANOQUESTFIT, INFN, and the University of Trieste (FRA 2016). They are indebted to Prof.~M.~Arndt and Prof.~H.~Ulbricht  for several stimulating and clarifying discussions. They also thank M.~Carlesso for preparing Fig.~2.

\section*{Appendix}

\subsection*{A1: $F$ and $D$ functions}
We present the analytical expressions of the $F$ functions in Eq.~\eqref{eq:1}~\cite{matterwave}. For the CSL model we have: 
\begin{equation}
\begin{split}
F_{\text{\tiny CSL}}(k,q,t)= \exp\bigg[&-\lambda \frac{m^2}{m_0^2} t \\
&\times\left(1-\frac{1}{t}\int_0^t d\tau e^{-\frac{1}{4r_C^2}(q-\frac{k\tau}{m})^2} \right) \bigg],
\end{split}
\end{equation}
where $m$ is the mass of the system, $m_0$ a reference (the nucleon's) mass, and $\lambda$ and $r_C$ are the CSL parameters previously introduced. 

For the dCSL model instead it takes the form:
\begin{equation}
\begin{split}
&F_{\text{\tiny dCSL}}(k,q,t)= \exp\Bigg[-\lambda \frac{m^2}{m_0^2} t \\ 
&\times \Big(1-\frac{1}{t} \int_0^t d\tau e^{-\frac{k^2r_C^2 k_T^2 }{\hbar^2} 
-\frac{(-\frac{k\tau}{m}+q)^2}{4r_C^2 (1+k_T)^2}}
e^{\frac{i(-k\tau+m q) 2k_T u_x }{\hbar(1+k_T)}}  \Big) \Bigg],
\end{split}
\end{equation}
where $k_T=\frac{\hbar^2}{8 m r_C^2 k_B T}$, $k_B$ is Boltzmann constant, $T$ the temperature the system thermalizes to, and $u_x$ is the $x$ component of the relative velocity between the noise field and the system. 

Finally, for the cCSL model it reads: 
\begin{equation}
\begin{split}
F_{\text{\tiny cCSL}}(k,q,t)= &F_{\text{\tiny CSL}}(k,q,t)\\
&\times\exp\Bigg[\frac{\lambda \bar{\tau}}{2}  \left(e^{-\frac{(q-\frac{k t}{m})^2}{4r_C^2}} - e^{-\frac{q^2}{4r_C^2}}\right) \Bigg],
\end{split}
\end{equation}
where $\bar{\tau}=\int_0^t s f(s) ds$, with $f(s)$ the  temporal correlation function of the noise. See Appendix A2 for a summary of the limits of validity of these expressions.

\begin{figure}[!t]
\includegraphics[width=0.75\columnwidth]{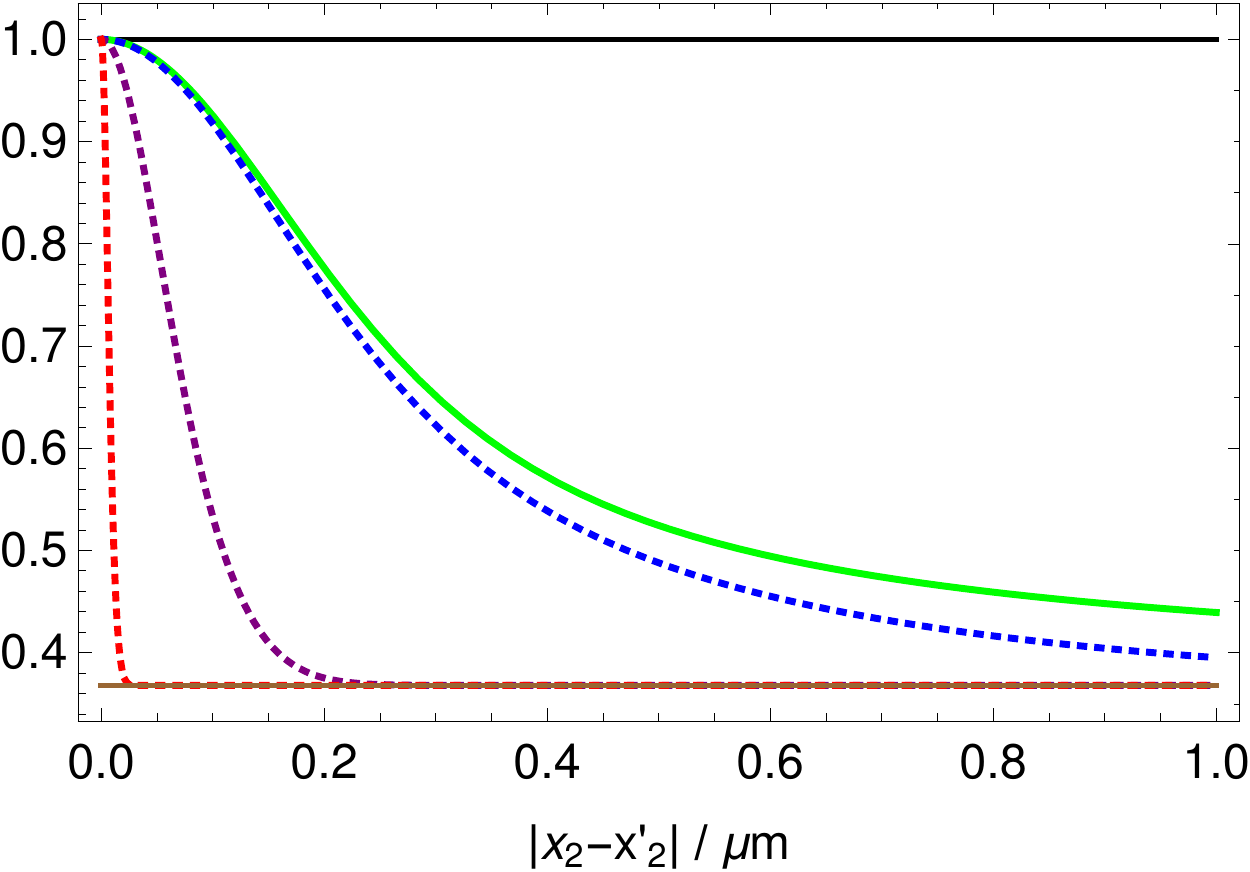}
\caption{Plot of the $D$ functions, quantifying the collapse effect, given by Eq.~\eqref{eq:3}. The black solid line represents the quantum mechanical function ($D=1$), the green solid line represents the $D$ function for the CSL and cCSL with $\Omega \gg 10^{13}\text{Hz}$, as well as the dCSL model with high noise temperature $T$ and low average $x$-axis noise-field velocity $|u_x|$ (see main text and Appendix A2 for details). The dashed lines represent the dCSL model with noise temperatures (average $x$-axis noise-field velocity) $T=10^{-8}\text{K}$ ($|u_x|\simeq 2\times 10^{4}\text{ms}^{-1}$), $T=10^{-9}\text{K}$ ($|u_x| \simeq 10^{5}\text{ms}^{-1}$)  and $T=10^{-10}\text{K}$ ($|u_x| \simeq 10^{6}\text{ms}^{-1}$) denoted by the color blue, purple and red, respectively. The solid brown line represents the asymptotic value of the $D$ functions for all the considered collapse models. The plot is obtained with typical flight times $t_1=t_2=1\text{ms}$ and distances $L_1=L_2=0.1\text{m}$ as in \cite{C3CP51500A}, with the usual value of $r_C=100\text{nm}$ and an exaggerated value $\lambda=500 \text{s}^{-1}$, to stress the different behaviour of collapse models with respect to ordinary quantum mechanics.}
\label{Fig:2}
\end{figure}
The corresponding $D$ functions are depicted graphically in Fig.~\ref{Fig:2}. We see that all functions share two common features: the initial value at $x=0$ is $1$ and the asymptotic value at $x\rightarrow \infty$ is given by $e^{-\lambda (t_1+t_2)}$, where $t_1$, $t_2$ denote the times of flight from the source to the grating and from the grating to the detector, respectively.

\subsection*{A2: Limits of validity of cCSL and dCSL bounds from matter-wave interferometry}
We report the limits of validity of the cCSL and dCSL bounds from the matter-wave inteferometry~\cite{matterwave}. 

We first discuss the limits of validity of the cCSL bounds, where the non-Markovian effects are parametrized by the correlation time $\tau_{C}$ (see main text). We can make a rough estimate for the maximum value of $\tau_{C}$ using a semi-classical calculation:
\begin{equation}
\langle\frac{{\hat{\bm p}}^{2}}{2m}\rangle\frac{\tau_{C}}{\hbar}\ll1,
\end{equation}
where $m$ is the mass of the molecule.  
Heuristically, this condition can be motivated by requiring that the evolution of the system is negligible on the time-scale $\tau_{C}$ of non-Markovian effects: 
$\hat{U}=\exp(-\frac{i}{\hbar}\frac{\hat{\bm p}^{2}}{2m}\tau_{C})\approx \hat{\mathbb{I}}$.
The typical temperature of the system is $T\approx10^{2}-10^{3}K$. Thus based on the equipartition theorem we replace $\langle{\hat{{\bm p}}}^{2}/2m\rangle$ by $k_{B}T$, which gives the condition $\tau_{c}\ll10^{-13}s$. This gives us a corresponding minimum frequency cut-off $\Omega\gg10^4\text{GHz}$ for the Fourier transform of the correlation function.

\begin{figure}[!t]
	\includegraphics[height=6cm]{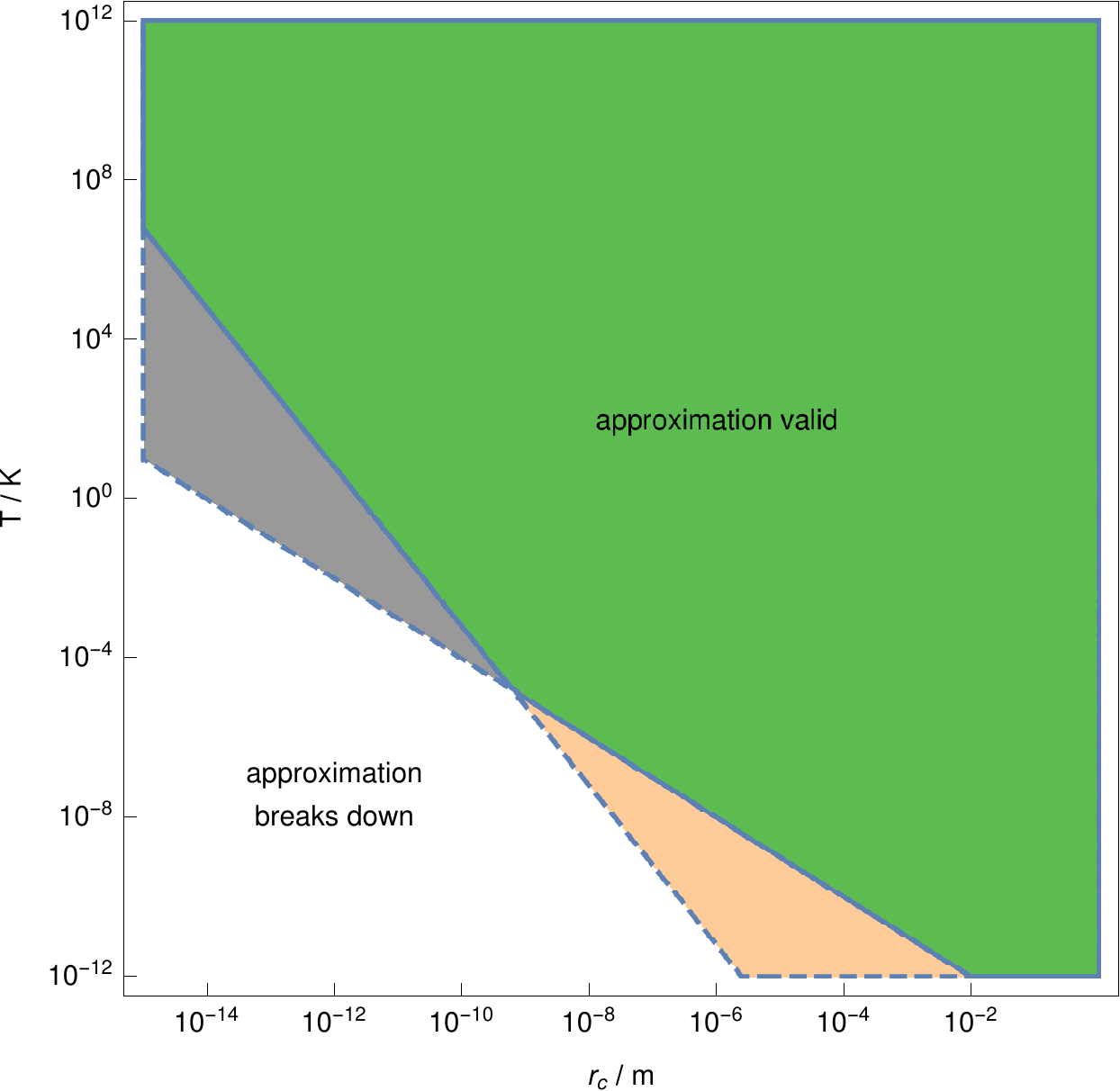}
	\caption{ Graphical depiction of the first two conditions given in Eq.~(\ref{eq:condtion}). Both conditions are satisfied in the green region. The first and second conditions are satisfied individually also in the orange  and gray regions, respectively.}
	\label{Fig:a1}
\end{figure}

\begin{figure}[!t]
  \includegraphics[height=6cm]{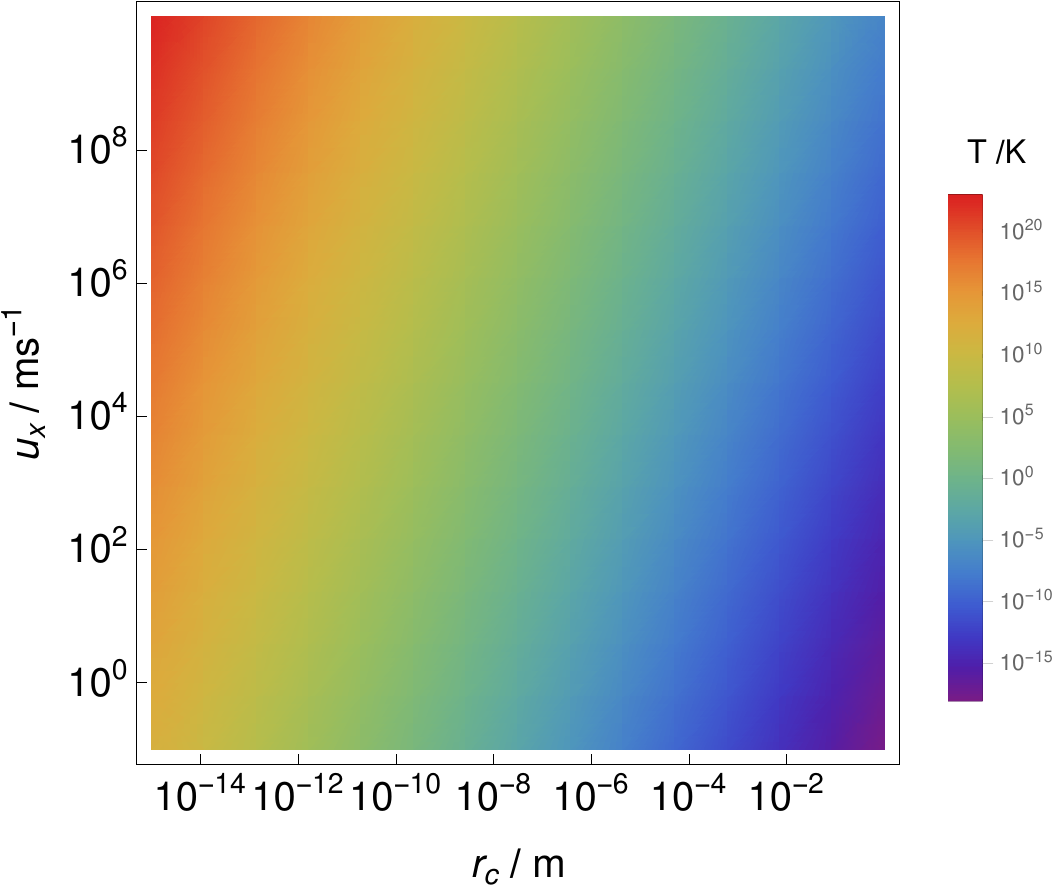}
  \caption{ Graphical depiction of the third condition given in Eq. (\ref{eq:condtion}). The color indicates the minimum temperature $T$, for a given value of $r_{C}$ and $u_x$, such that the third condition given in Eq.~(\ref{eq:condtion}) is satisfied.}
  \label{Fig:a2}
\end{figure}

We next discuss the limits of validity of the dCSL bounds, where the dissipative effects are parametrized by the temperature $T$ and by the boost parameter $u_x$ along the $x$ axis (see main text). We have three conditions:
\begin{equation}
\frac{\hbar^2}{mr_C^2}, \, \frac{\hbar \Delta x}{r_C t}, \,  \frac{\hbar \Delta x u_x }{r_C^2} \ll 8 k_B T  
\label{eq:condtion}
\end{equation}
where $k_{B}$ is Boltzmann's constant, $m$ is the mass of the system, $\Delta x$ is the size of the system along the $x$ axis and $t$ is the time of flight. These conditions, which quantify how close we are to the non-dissipative regime, have a straighforward heuristic motivatation. The first condition quantifies the overall degree of dissipation which, close to the non-dissipative regime, is proportional to $\propto T^{-1}$~\cite{gasbarri2017general}, while the second and third condition require that the characteristic velocities of the system and noise, i.e. $\Delta x/t$ and $u_x$, respectively, are not too large. The latter two, if violated, would again lead to strong dissipation. The first two conditions are depicted in Fig.~\ref{Fig:a1}, while the third condition is depicted in Fig.~\ref{Fig:a2}.

\subsection*{A3: Colored and dissipative CSL}\label{app:Color_diss}

We first construct the dCSL master equation. The assumption of (i) translational covariance implies that the master equation has the Holevo structure \cite{holevo1993note,holevo1993conservativity,holevo1995translation,holevo1996covariant} (in the Schr\"{o}dinger picture):
\begin{alignat}{3}
\frac{\partial\hat{\rho}_{t}}{\partial t}=&-\frac{i}{\hbar}[\hat{H},\hat{\rho}_{t}]
&&\nonumber\\
&+\int_{\mathbb{R}^3} d\bm{Q}\bigg(
&& e^{\frac{i}{\hbar}\bm{Q}\cdot\hat{\bm{x}}}J(\hat{\bm{p}},\bm{Q})\hat{\rho}_{t}J^{\dagger}(\hat{\bm{p}},\bm{Q})e^{-\frac{i}{\hbar}\bm{Q}\cdot\hat{\bm{x}}} \nonumber\\
& &&-\frac{1}{2}\{J^{\dagger}(\hat{\bm{p}},\bm{Q})J(\hat{\bm{p}},\bm{Q}),\hat{\rho}_{t}\}\bigg).
\label{eq:rho_sch}
\end{alignat}
We assume $J$ is a (iii) Gaussian in $\hat{\bm{p}}$ and $\bm{Q}$:
\begin{alignat}{1}
J(\hat{\bm{p}},\bm{Q}) & =N\exp(-(a_{1}\bm{Q}+a_{2}\hat{\bm{p}})^{2}),\label{eq:JoverDRe2}
\end{alignat}
where $a_{1}$, $a_{2}$ are real-valued parameters and $N$ is a normalization constant.

We now impose (ii) that the Gibbs state ($\hat{\rho}_{\text{asm}}$)
defined in Eq.~(\ref{eq:rho_asm}) is a stationary state of Eq. (\ref{eq:rho_sch}), which yields the condition:
\begin{equation}
\begin{split}
\int_{\mathbb{R}^3} d\bm{Q}e^{-2(a_{1}\bm{Q}+a_{2}(\hat{\bm{p}}-\bm{Q}))^{2}}e^{-\beta(\hat{\bm{p}}-\bm{Q})^{2}/(2m)}\nonumber\\
=\int_{\mathbb{R}^3} d\bm{Q}e^{-2(a_{1}\bm{Q}+a_{2}\hat{\bm{p}})^{2}}e^{-\beta\hat{\bm{p}}{}^{2}/(2m)}.
\end{split}
\label{eq:a1a2}
\end{equation}
The above equation is satisfied if
\begin{equation}
\begin{split}
-2(a_{1}\bm{Q}+a_{2}(\hat{\bm{p}}-\bm{Q}))^{2}-\beta(\hat{\bm{p}}-\bm{Q})^{2}/(2m) \nonumber\\
=-2(a_{1}(\bm{Q}+\bm{b})+a_{2}\hat{\bm{p}})^{2}-\beta\hat{\bm{p}}{}^{2}/(2m),
\end{split}
\end{equation}
where $\bm{b}$ is a vector with $\mathbb{Re}$-valued components. Looking at each power in $\bm{Q}$, we obtain three
conditions:

\begin{alignat}{1}
\bm{Q}^{0}: & \:a_{1}^{2}\bm{b}^{2}+2a_{1}a_{2}\bm{b}\cdot\hat{\bm{p}}=0,\label{eq:Q0}\\
\bm{Q}^{1}: & \:2a_{2}^{2}\hat{\bm{p}}+2a_{1}^{2}\bm{b}+\frac{\beta}{2m}\hat{\bm{p}}=0,\label{eq:Q1}\\
\bm{Q}^{2}: & \:-2a_{2}^{2}+4a_{1}a_{2}-\frac{\beta}{2m}=0.\label{eq:Q2}
\end{alignat}
From Eq. (\ref{eq:Q0}) we obtain two solutions:
\begin{equation}
\bm{b}=-2(a_{2}/a_{1})\hat{\bm{p}}\label{eq:b}
\end{equation}
and $\bm{b}=0$. However, we do not consider $\bm{b}=0$ as it leads to an imaginary value for $a_{2}$ (see Eq. (\ref{eq:Q1})), in contradiction with our Ansatz in Eq. (\ref{eq:JoverDRe2}). The remaining two Eqs.~(\ref{eq:Q1}), (\ref{eq:Q2}) are not independent: we can write $a_{1}$, $a_{2}$ as functions of a free parameter, which we denote by $r_{C}$. Specifically, we write them, for later convenience, as:
\begin{alignat}{1}
a_{1} & =(1+k_{T})/(\sqrt{2}\hbar/r_{C}),\label{eq:a1}\\
a_{2} & =2k_{T}/(\sqrt{2}\hbar/r_{C}),\label{eq:a2}
\end{alignat}
where 
\begin{equation}
k_{T}=\frac{\hbar^{2}}{8mr_{C}^{2}k_{B}T}.
\end{equation}
To summarize we have the following operator
\begin{equation}
J(\hat{\bm{p}},\bm{Q})=N\exp(-(r_{C}^{2}/(2\hbar^{2}))((1+k_{T})\bm{Q}+2k_{T}\hat{\bm{p}})^{2}),
\label{eq:JoperatorRe2}
\end{equation}
where the overall normalization, chosen as in \cite{Smirne:2014paa},
is determined by the free parameter $\lambda$:
\begin{equation}
N=\sqrt{\lambda\frac{m^{2}}{m_{0}^{2}}\left(\frac{r_{C}}{\sqrt{\pi}\hbar}\right)^{3}},
\end{equation}
In particular, $\lambda$ can be interpreted as the collapse rate of a reference object with mass $m_0$ and $r_C$ as a characteristic localization lenght (see main text). Moreover, the quadratic dependence on the mass $m$ of a single particle, \textit{i.e.} $m^{2}/m_{0}^{2}$, is motivated by the amplification mechanism: the localization rate of a point-like system with $N$ particles is amplified by a factor $N^2$~\cite{collapse_review1}.

The dCSL model described by Eq.~\eqref{eq:rho_sch}, can be easily generalized to account for the non-white spectrum of a more physical noise. Specifically, we consider the framework of completely positive (CP) non-Markovian Gaussian maps~\cite{diosi2014general} that generalizes the dynamics of Lindblad-type master equations such as the one in Eq.~\eqref{eq:rho_sch}. Imposing the request of translational covariance on a general CP Gaussian map one obtains the dynamical map in Eq.~\eqref{eq:translationC1}~\cite{gasbarri2017general}. This map is characterized by the operators $J$ and by the temporal correlation function $G$. By choosing the operators $J$ obtained in Eq.~\eqref{eq:JoperatorRe2} also for the non-Markovian dynamics in Eq.~\eqref{eq:translationC1} we obtain the cdCSL dynamical map discussed in the main text. 

Conversely, the dCSL model can be seen as the limit of the cdCSL model, when the evolution time $t$ is long and the correlation time $\tau_C$ is short. Loosely speaking, in this regime a non-Markovian dynamics can be approximately described by a Markovian dynamics. For more details about this regime see the corresponding analysis for CP non-Markovian Gaussian maps, contained in the supplemental material of Ref.~\cite{gasbarri2017general}.


\bibliography{numoqm}
\end{document}